\documentclass[twocolumn]{aastex631}
\usepackage{blindtext}

\shorttitle{Revisiting BD-06~1339b}
\shortauthors{Emilie R. Simpson et al.}

\begin{document}

\title{Revisiting BD-06~1339b: A Likely False Positive Caused by Stellar Activity}

\author[0000-0003-0447-9867]{Emilie R. Simpson}
\affiliation{Department of Earth and Planetary Sciences, University of California, Riverside, CA 92521, USA}
\email{esimp005@ucr.edu}
  
\author[0000-0002-3551-279X]{Tara Fetherolf}
\altaffiliation{UC Chancellor's Fellow}
\affiliation{Department of Earth and Planetary Sciences, University of California, Riverside, CA 92521, USA}

\author[0000-0002-7084-0529]{Stephen R. Kane}
\affiliation{Department of Earth and Planetary Sciences, University of California, Riverside, CA 92521, USA}

\author[0000-0002-4860-7667]{Zhexing Li}
\affiliation{Department of Earth and Planetary Sciences, University of California, Riverside, CA 92521, USA}

\author[0000-0002-3827-8417]{Joshua Pepper}
\affiliation{Department of Physics, Lehigh University, 16 Memorial Drive East, Bethlehem, PA 18015, USA}

\author[0000-0003-4603-556X]{Teo Mo\v{c}nik}
\affiliation{Gemini Observatory/NSF's NOIRLab, 670 N. A'ohoku Place, Hilo, HI 96720, USA} 

  
\begin{abstract}

As long as astronomers have searched for exoplanets, the intrinsic variability of host stars has interfered with the ability to reliably detect and confirm exoplanets.  One particular source of false positives is the presence of stellar magnetic or chromospheric activity that can mimic the radial-velocity reflex motion of a planet. Here we present the results of a photometric data analysis for the known planet hosting star, BD-06~1339, observed by the Transiting Exoplanet Survey Satellite (TESS) during Sector 6 at 2 minute cadence. We discuss evidence that suggests the observed 3.9 day periodic radial velocity signature may be caused by stellar activity rather than a planetary companion, since variability detected in the photometric data are consistent with the periodic signal. We conclude that the previously reported planetary signature is likely the result of a false positive signal resulting from stellar activity, and discuss the need for more data to confirm this conclusion. 

\end{abstract}


\section{Introduction}
\label{intro}

Exoplanets discoveries thus far have been dominated by indirect techniques, mostly due to the success of the radial velocity (RV) and transit techniques. Prior to the discoveries of the Kepler mission \citep{borucki2010a,borucki2016}, the majority of exoplanets were discovered using the RV method \citep{butler2006, schneider2011}, with a growing number of ground-based transit discoveries \citep{konacki2003, alonso2004, bakos2007, kane2008a}. Indirect detection techniques rely on a detailed characterization of the host star, since the properties of the host star determine the extracted planetary parameters \citep{seager2003,vanbelle2009a}. Of particular importance is the effect of stellar activity, since this can severely limit the detection of exoplanets around active stars \citep{desort2007,aigrain2012,zellem2017}, and can even result in false-positive detections, whereby stellar activity cycles can masquerade as exoplanet signatures \citep{nava2020}. Indeed, there have been numerous instances of exoplanet claims using the RV method that were later determined to be the result of stellar activity \citep{henry2002, robertson2014, robertson2015, kane2016a}. This potential confusion may be mitigated in certain cases by utilizing precision photometry for known exoplanet hosts \citep{kane2009c}, such as data acquired by transit surveys. A transit detection of an RV planet can provide confirmation of the planet, as well as provide an additional means to disentangle stellar variability and planetary signatures \citep{boisse2011, diaz2018b}. The Transiting Exoplanet Survey Satellite (TESS) \citep{ricker2015} provides an invaluable photometric data source for known exoplanet hosts \citep{kane2021b} since it is monitoring most of the sky, and is especially well-suited for observing the bright host stars typical of RV exoplanet searches \citep{fischer2016}.

Stellar activity has long been known to affect and sometimes limit RV exoplanet searches \citep{saar1997b}, and can particularly impact detection of planets within the Habitable Zone \citep{vanderburg2016b}. Photometric monitoring of known host stars has been used in numerous cases to determine the effects of their variability on planetary signatures, such as for HD~63454 \citep{kane2011e} and HD~192263 \citep{dragomir2012a}. Another example of a host star exhibiting significant stellar variability is the case of BD-06~1339, which was discovered to host planets by \citet{locurto2013} using data from the High Accuracy Radial velocity Planet Searcher (HARPS) spectrograph \citep{pepe2000}. These observations revealed two planetary signatures with orbital periods of 3.87 days and 125.94 days, with minimum planetary masses of 0.027 and 0.17 $M_J$, respectively. However, photometry of sufficient precision, cadence, and duration was not available in order to confirm a transit signature.

Here, we present an investigation into the BD-06~1339b planetary signature by analyzing the associated TESS photometry and re-analyzing the existing HARPS RV data. In Section~\ref{system}, we discuss the properties of the system, including the stellar parameters, and the possible planets within the system. Section~\ref{data} describes the data analysis for the system, where the data sources are comprised of HARPS RV data and the precision photometry from TESS. Section~\ref{false} combines these results to present an argument that the RV variations originally detected could alternatively be consistent with the intrinsic variability of the host star. We provide concluding remarks in Section~\ref{conclusions}, and outline how the photometric capabilities from TESS not only serve to discover new planets, but also have considerable utility in testing known exoplanet hypotheses.


\section{System Properties}
\label{system}

BD-06~1339 (HIP~27803, GJ~221, TIC~66914642) is a relatively bright high proper-motion star located at a distance of 20.27~pcs \citep{brown2018,brown2021}. According to \citet{locurto2013}, BD-06~1339 is a late-type dwarf star, with a spectral classification of K7V/M0V and an age similar to that of the Sun. The star has an effective temperature of 4324~K, a V magnitude of 9.70, and a stellar mass of 0.7~$M_\odot$. Initial spectroscopic analyses were performed in 1996 for the Palomar/MSU Nearby Star Spectroscopic Survey \citep{hawley1996} among previously reported variable stars. A further survey of chromospheric activity among cool stars by \citet{borosaikia2018} found that BD-06~1339 is moderately active, with an activity index of $\log R'_{HK} = -4.71$. Such magnetic activity is prevalent in later stellar spectral types \citep{mcquillan2012}, lending to the stellar activity of interest for this study. 

The host star is currently reported to have two companions, BD-06~1339b and BD-06~1339c, both of planetary mass and discovered via the RV technique \citep{locurto2013}. Though discovered simultaneously, their properties differ greatly; BD-06~1339b has a minimum mass of 8.5~$M_\oplus$ and orbits its host star in 3.873 days at a semi-major axis of 0.0428~AU. Its sibling, BD-06~1339c, has a minimum mass of 53~$M_\oplus$, has an orbital period 125.94 days at a semi-major axis of 0.435~AU. The \citet{locurto2013} analysis of the RV data for BD-06~1339b adopts a fixed circular orbit ($e = 0$) for the b planet, and derives an eccentricity of 0.31 for the c planet. \citet{tuomi2014} conducted a statistical reanalysis of the RVs for BD-06 1339, which we further investigate in Section~\ref{rv}.


\section{Data Analysis}
\label{data}

The motivation for re-analyzing BD-06~1339b stems from a broad stellar variability analysis of stars observed during the TESS primary mission at 2-min cadence (Fetherolf et al. in prep.), and a further investigation into the stellar variability of known exoplanet host stars (Simpson et al. in prep.). The broad stellar variability analysis by Fetherolf et al. (in prep.) searches for periodic photometric modulations on timescales up to the duration of a single orbit of the TESS spacecraft (0.01--13~days), during which TESS obtained continuous observations. The $\sim$700 exoplanet host stars that were selected for the follow-up variability analysis by Simpson et al. (in prep.) includes planets with orbital periods shorter than 13 days that were discovered by either their RV or transit signatures. Since Kepler exoplanet host stars are typically faint and not ideal for RV follow-up observations, they are not included in the stellar variability analysis of known exoplanet host stars. In addition to possible transit events or variations due to stellar activity, some of these planets may also exhibit interactions with their host stars, such as phase variations or star-surface irregularities.

The full TESS light curve, Lomb-Scargle (L-S) periodogram, and light curve that was phase-folded on the most significant photometric variability signature were each visually inspected for the $\sim$700 known exoplanet host stars. The photometric periodicity was determined to be significantly variable if both the normalized and phase-folded light curves displayed sinusoidal behavior that did not align with known spacecraft systematics (i.e., momentum dumps), and if the periodogram maximum exhibited an isolated peak with at least 0.001 normalized power that also exceeded the 0.01 false alarm probability level. For each known exoplanet, the extracted photometric variability period was compared to their orbital period, as reported by either the TESS Objects of Interest (TOI) catalog \citep{guerrero2021} or by cross-referencing the target in the NASA Exoplanet Archive \citep{nea}. Close-period matches between the photometric variability and the planetary orbital period were defined as being within 5--10\% of each other. Out of the $\sim$700 targets subjected to the visual analysis, approximately 180 systems displayed prominent photometric variable behavior, close-period matches, or both.

\begin{figure*}
    \centering
    \includegraphics[width=\textwidth]{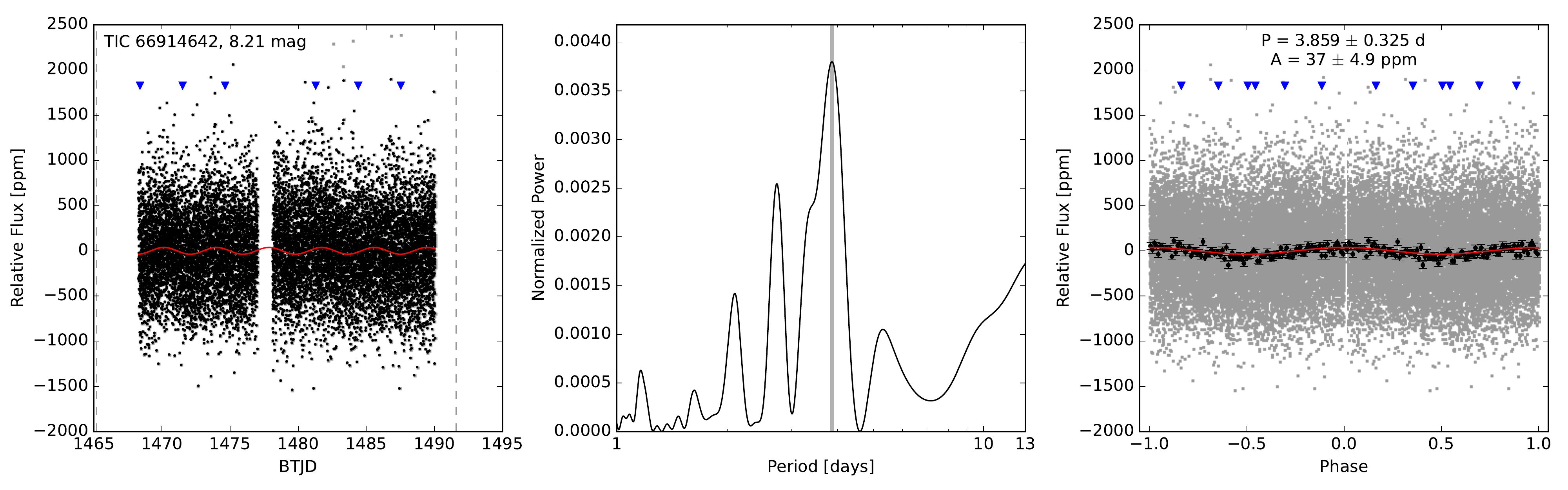}
    \caption{The TESS light curve (left), periodogram (center), and phase-folded light curve (right) of BD-06~1339. The red curve represents a sinusoidal model fit representing the light curve variability. The blue triangles indicate timings of spacecraft thruster firings (i.e., momentum dumps). This system stood out in our visual analysis because of its pronounced sinusoidal behavior and the strong single peak in the periodogram, which indicates a strong variability signal. The detected variability periodicity (3.859\,days) is only 0.03\% away from the reported orbital period for BD-06~1339b (3.873\,days).}
    \label{fig:variability}
\end{figure*}

BD-06~1339b was among the set of targets that matched these criteria, and the resulting TESS light curve, periodogram, and phase-folded light curve are shown in Figure~\ref{fig:variability} (see also Section~\ref{phot}). In this paper we revisit the analysis of the BD-06~1339 system by including the TESS photometry that was unavailable at the time of previous studies. In Section~\ref{rv} we summarize our re-analysis of the RVs using the data provided by the updated HARPS reduction pipeline \citep{trifonov2020}. We then discuss our in-depth analysis of the TESS photometry in Section~\ref{phot}, where we search for the presence of planetary transits, atmospheric variations, and stellar activity.

\subsection{Spectroscopic Analysis}
\label{rv}

\citet{locurto2013} verified the planets orbiting BD-06~1339 by requiring the normalized fourier power of the L-S periodogram of the RV time series to have a false alarm probability of $<{10^{-4}}$. The $\log R'_{HK}$  activity index was considered poor quality, and therefore was not utilized in the overall analysis. BD-06 1339b was barely discernible within the RV signal of BD-06 1339c in this stage of analysis, existing initially as additional variations. To verify the planetary nature of this signal, the founding team cut the datasets in half to exclude long-term trends. These variations then increased in strength throughout the observation period as the Ca II H re-emission decreased. The discovery team was unable to determine any other longer-term trends due to their limited window of 102 observations over 8 years. They determined that BD-06 1339 b was an educational case of how planets can hide within the activity of variable stars.

To further analyze the BD-06 1339 system, \citep{tuomi2014} implemented a more meticulous probability check involving an independent statistical method of subsequent samplings and the utilization of log-Bayesian evidence ratios. The Bayesian analysis used by \citet{tuomi2014} evaluated the RV time series as if they were observed in real time. At each iteration, a ``new'' RV measurement was added to the dataset from which the best-fit system parameters were determined. In this case, they utilized HARPS and Planet Finder Spectrograph (PFS; \citealp{crane2010}) velocities in their credibility tests. HARPS found the planetary signals of the two original targets and consistent with each other. PFS could not discern any signals previously found by \citet{locurto2013}. The system itself was not explicitly observed for this publication, instead relying on the previous data available at the time. Their results focused on the discovery of a third d planet at a $\sim$400\,day orbital period based on a statistical probability, and they considered BD-06~1339b as a confirmed planet. 

The observations of BD-06~1339 were acquired by the HARPS team and originally published by \citep{locurto2013}. The data has since been re-reduced and includes corrections of several systematics within the observations \citep{trifonov2020}. With the improved precision and availability of the data, a re-analysis could derive the orbital parameters of the known companions in the system with better precision and potentially reveal smaller signals that were previously unreported before the re-reduction of the RVs. 

We performed a re-analysis of the RVs for BD-06~1339 using the re-reduced data published by \citet{trifonov2020}. We first ran an RV Keplerian periodogram on the dataset to search for significant signals using \texttt{RVSearch} \citep{rosenthal2021}. The \texttt{RVSearch} algorithm iteratively searches for periodic signals present in the dataset and calculates the change in the Bayesian Information Criterion ($\Delta$BIC) between the model at the current grid and the best fit model based on the goodness of the fit. The result of the search would yield signals that are of planetary origin as well as those that are due to stellar activity. We adopted signals returned by \texttt{RVSearch} if they peak above the 0.1\% false alarm probability level. The search returned two significant signals, one at 125 days and another at 3.9 days. This is consistent with the results from \citet{locurto2013}.

We then used the RV modeling toolkit \texttt{RadVel} \citep{fulton2018a} to fully explore the orbital parameters of these two signals and to assess their associated uncertainties. We provided the orbital parameter initial guesses for the two signals using the values returned by \texttt{RVSearch} and allowed all parameters to vary, including an RV vertical offset, RV jitter, and a linear trend. We fit the data with maximum a posteriori estimation and explored the posteriors of the parameters though Markov Chain Monte Carlo (MCMC). The MCMC exploration successfully converged and we show the results in Table~\ref{tab:param} below. 

Orbital parameters of the two signals are mostly consistent with those reported by \citet{locurto2013}, except that the orbit of the c planet is preferred to be nearly circular ($e_c\sim0.09$) instead of a mildly eccentric ($e\sim0.31$), as proposed by \citep{locurto2013}. In addition, there appears to be a significant linear trend ($\sim7\sigma$) present in the data that could be indicative of an additional long orbital period massive companion orbiting in the outer regime of this system. Both the linear trend and two circular orbits model are supported by Bayesian model comparisons. The RV signature for BD-06~1339b is shown in left panel of Figure~\ref{fig:RvvsFlux}, where the contribution from the c planet has been removed. The results of this latest RV re-analysis are consistent within the uncertainties of the original analysis performed by \citet{locurto2013}. 

\begin{deluxetable}{lcc}[tbp]
    \tablecaption{Updated RV System Parameters of BD-06~1339.
    \label{tab:param}}
    \tablehead{
        \colhead{Parameters} &
        \colhead{b} &
        \colhead{c}}
    \startdata
    $P$ (days) & $3.87302^{+0.00036}_{-0.00033}$ & $125.49\pm0.13$  \\ 
    $Tc$ (BJD) & $2455000.91^{+0.21}_{-0.17}$ & $2455279.6^{+2.0}_{-1.8}$    \\ 
    $Tp$ (BJD) & $2455001.65^{+0.39}_{-0.62}$ & $2455285^{+16}_{-14}$  \\ 
    $e$ & $0.22^{+0.16}_{-0.13}$ & $0.089^{+0.054}_{-0.052}$\\ 
    $\omega$ (deg) & $181.23^{+38.39}_{-55.00}$ & $110.58^{+49.27}_{-38.39}$ \\ 
    $K$ (m s$^{-1}$) & $3.47^{+0.52}_{-0.49}$ & $8.32^{+0.46}_{-0.47}$ \\ 
    $M_p$ ($M_{\rm E}$) & $6.45^{+1.0}_{-0.98} $ & $50.9^{+4.5}_{-4.4}$ \\ 
    $a$ (au) & $0.0429^{+0.0014}_{-0.0015}$ & $0.436^{+0.014}_{-0.015}$ \\ 
    \enddata
    \tablecomments{$\omega$ values are those of the star, not of the planet. The RV fit includes a linear trend of $\dot{\gamma}$ = $-0.00239^{+0.00032}_{-0.00033}$ m s$^{-1}$ d$^{-1}$.}
\end{deluxetable}


\subsection{Photometric Analysis}
\label{phot}

\citet{gillon2017c} used the Warm mode of the Spitzer mission to search for transits of 24 low-mass planets (all single planet systems) discovered through the RV method, including BD-06~1339b. The Spitzer photometry found no reliable transits for 19 of the 24 planets, including BD-06 1339b. Specifically, BD-06~1339b was found to not display a transit within the observation window, although the photometry did not cover approximately 20\% of the possible transit window. Since then, TESS observed BD-06~1339 at 2-min cadence nearly continuously during the observations of Sector 6. In this section, we use to the TESS photometry to search for transits by either the b or c planets and atmospheric phase variations caused by the b planet. 

\begin{figure*}
    \plottwo{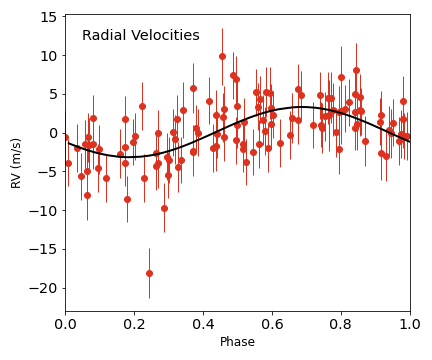}{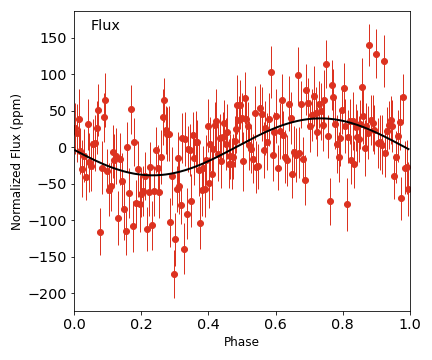}
    \caption{Left: The RV signature of BD-06 1339b (c planet's signal removed). Right: The TESS photometry phase-folded on the orbit of BD-06~1339b. The black curves represent sinusoidal fits to the data.}
    \label{fig:RvvsFlux}
\end{figure*}

\begin{figure}
    \includegraphics[width=0.5\textwidth]{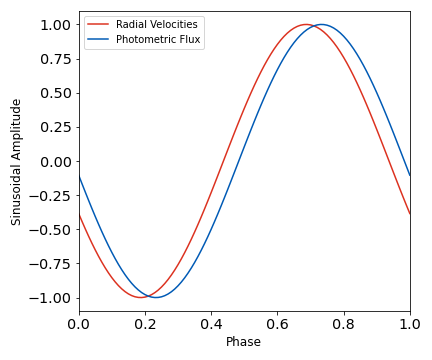}
    \caption{The sinusoidal fits to the RV (red) and flux (blue) curves from Figure~\ref{fig:RvvsFlux}, but with the amplitudes normalized to unity. A strong correlation in the phase offset can be seen around 0.7 phase.}
    \label{fig:Correlate}. 
\end{figure}

BD-06~1339 was observed during TESS Sector 6 (2018~Dec~11--2019~Jan~07) at 2-min cadence and TESS Sector 33 (2020~Dec~17--2021~Jan~13) at 30-min cadence. The TESS light curves and full-frame images are publicly available through the Mikulski Archive for Space Telescopes\footnote{\url{https://archive.stsci.edu/}} (MAST). Since the anticipated transit of BD-06~1339b is on the order of $\sim$2\,hr, we elect to only use the 2-min cadence light curve from the Sector 6 observations. We use the original data release of the pre-search data conditioning (PDC) light curve that was processed by the Science Processing Operations Center (SPOC) pipeline \citep{Jenkins2016} and additionally remove any observations denoted with poor quality flags or that are 5$\sigma$ outliers. The L-S periodogram \citep{Lomb1976, Scargle1982} is then computed on the BD-06~1339 light curve using an even frequency spacing of 1.35\,min$^{-1}$, where we find a maximum normalized power of $\sim$0.0038 at $3.859\pm0.325$\,days. The 0.01 false alarm probability level for the periodogram of the BD-06~1339 light curve corresponds to 0.0012 normalized power, with the peak of the periodogram having a $\ll$10$^{-4}$ false alarm probability.

Our L-S periodogram analysis of the TESS light curve reveals a sinusoidal periodicity that is consistent with the orbital period of the b planet ($3.8728\pm0.0004$~days) within their uncertainties (see Figure~\ref{fig:variability}). A planet's orbital period may be extracted from a periodogram analysis if transit events are not properly removed from the observed light curve. However, we do not observe transit events by either the b or c planets in the TESS photometry (see Figure~\ref{fig:RvvsFlux}), which is consistent with the findings of \citep{gillon2017c}. A significant sinusoidal amplitude could also indicate the presence of a planet-induced photometric phase curve caused by its day-side reflection or excess thermal emission. If the phase curve is caused by the day-side reflection of the planet, then the maximum brightness of the phase-folded light curve is expected to peak at 0.5 phase when we see the greatest area of the planet illuminated from our point of view. Alternatively, atmospheric winds that redistribute heat from the day to night-side could cause the hottest region of the atmosphere to be shifted eastwards from the sub-stellar point, such that the phase-folded light curve peaks prior to 0.5 phase \citep[e.g.,][]{Showman2013, Heng2015}. 

We use the measured time of conjunction (i.e., expected transit time) from the RV analysis to assess both the shape and phase of maximum amplitude of the TESS phase-folded light curve for BD-06~1339b. The full phase curve is fit using a double harmonic sinusoidal function, which allows for modulations caused by Doppler boosting and ellipsoidal variations in addition to the reflection caused by the day-side of the planet \citep[see][]{Shporer2017a}. The first cosine harmonic component represents the modulations caused by day-side reflection or thermal emission, such that the maximum brightness occurs at 0.5 phase. The phase-folded light curve of BD-06~1339b exhibits a significant sinusoidal modulation of $\sim$40 ppm in the TESS photometry (see right panel of Figure~\ref{fig:RvvsFlux}) with a maximum brightness at the third quadrature of the b planet's orbital phase (0.73 phase). In addition to the maximum brightness being at a phase that is inconsistent with day-side reflection or thermal emission, the amplitude of the phase curve is $\sim$10 times greater than expected for such a small planet.\footnote{The day-side reflection modulations of 8.5\,$M_\oplus$ planet with an albedo of 0.3 are expected to be on the order of $\sim$2\,ppm.}



\section{False-Positive Planetary Signature?}
\label{false}

The results described above cast doubt on the planetary origin of the signal ascribed to BD-06~1339b. This target may, in fact, instead be a possible case for the stellar variability of the host star masquerading as a false positive. While it is not impossible for a system to exist in which a planet orbits at the same period as its host star's variability, a coincidence of 0.01 days between the two is highly unlikely. A visual comparison of the RVs and stellar flux of the host star is enough to raise some questions, but we must quantify our results. We further investigate the nature of BD-06~1339b by comparing the phase signature in the RVs and photometry, searching for correlations in the spectral activity indicators, and considering the likelihood of BD-06~1339 exhibiting periodic stellar activity at $\sim$3.9\,days. 

Figure~\ref{fig:RvvsFlux} shows the RV signature and the photometric variations in phase with the anticipated orbit of BD-06~1339b. We fit a simple sinusoidal function to each phase curve and find that the maximum of the RVs occurs at 0.69 phase with an amplitude\footnote{This amplitude is estimated assuming a simple sinusoidal function, and thus a zero eccentricity.} of 3.3\,m\,s$^{-1}$, and the maximum of the photometric flux occurs at 0.73 phase with an amplitude of 40\,ppm.  Interestingly, the RVs and the photometric variations peak at approximately the same phase. The correlation between these sinusoidal functions is further emphasized in Figure~\ref{fig:Correlate}, where the two functions are normalized by having their amplitudes set to unity. 

Clearly there is a very strong correlation between the RVs and the photometry, but they should instead be offset from each other in phase. If the photometric variations were caused by atmospheric reflection or thermal emission of BD-06 1339b, the photometric variations should peak at 0.5 phase or earlier due to winds \citep[e.g.,][]{Showman2013, Heng2015}. However, the observed phase offset is subject to uncertainties from the time of conjunction determined from the RVs (0.2\,days) and the time between the RV and TESS observations ($\sim$3500\,days). Propagating the time of conjunction, and thus phase offset, out to the time of the TESS observations results in an uncertainty of 0.5\,days (13\% of the orbital period), which could render the correlation in phase between the RVs and photometry as a coincidence.  

In addition to the photometry, we performed an analysis on all of the available RV activity spectral indicators provided by the HARPS RV database \citep{trifonov2020} to investigate whether any significant activity signals are consistent with the reported period for BD-06~1339b. We used a Generalized L-S periodogram (GLS; \citealp{zechmeister2009}) to search for periodicity in H$_{\alpha}$, chromatic index (CRX), differential line width (dLW) \citep{zechmeister2018}, as well as full-width-at-half-maximum (FWHM) and contrast of the cross correlation function (CCF). None of the aforementioned indicators returned significant signals above the 0.1\% false alarm probability level, except for dLW where a $\sim$ 270-day signal was detected just above the false alarm probability threshold and is possibly of stellar activity origin. 

We also investigated if there exists any correlation between the b planet's RV signal (after the removal of RV contributions from the c planet and the linear trend) and each one of the activity indicators using the Pearson correlation coefficient. Once again, only dLW returns a weak correlation of $\sim$ 0.25, while there is no correlation observed in any of the other activity indicators. While there is no peak in the dLW periodogram near $\sim$ 3--4 days, the correlation between the b planet's RV signature and dLW could be related to the 270-day signal. Overall, despite the strong indication from the photometry that the previously reported b signal could attributed to stellar activity, no significant correlations were found between the b planet's RVs and any of the spectral activity indicators, and no activity periods were detected near the b planet's orbital period.

This raises the question of how stellar variability can be selectively manifesting in the photometry, but not in the spectral lines of the host star. We investigate whether the signal observed in the BD-06~1339 light curve is typical for stars of similar spectral types. From the all-sky variability analysis, we searched for stars with effective temperatures between 4000--4500\,K, photometric variability periods of 3.5--4.0\,days, and stellar luminosities lower than 10\,$L_\odot$. We find $\sim$30 stars within this subgroup and, upon visual investigation, find that their light curves are similar in shape and amplitude to the variations observed for BD-06~1339 (see Figure~\ref{fig:variability}). Their light curve behavior proved to be comparable to what is observed in the BD-06~1339 light curve. Therefore, stellar activity is a potential explanation for the observed photometric variations.

We cannot pinpoint the physical mechanism behind the photometric variability, although our general understanding of stellar astrophysics suggests that it is related to magnetic activity in the star that produces spots and plages. The false alarm probability \citep{locurto2013} used to detect BD-06~1339b was based on a simple f-test, but recent work has shown that other statistical methods, such as the extreme value statistical distribution, may be more appropriate for applying to periodogram analyses \citep{suveges2014, vio2016, sulis2017, vio2019, delisle2020}. The close match between both the period and phase of the photometric variability and the RV variations suggests that both signals are produced by the same cause.  We therefore believe that the most likely explanation is that BD-06~1339b is a false positive and that the RV variations are not produced by a planetary companion of the star.


\section{Conclusions}
\label{conclusions}

We conducted a photometric analysis of targets with periodic modulations from the TESS primary mission (Fetherolf et al. in prep.; Simpson et al. in prep.) and determined that BD-06~1339b was considered a prime subject for further scrutiny. The similarity between the photometric variability periodicity of the TESS photometry for BD-06~1339 (3.859\,days) and the orbital period of the b planet (3.874\,days) prompted a rigorous re-examination of the spectroscopy and photometry for this target. We performed a re-analysis of the RVs obtained by HARPS and found an orbital solution that was consistent with the RV analysis performed by \citet{locurto2013}. An in-depth investigation of the photometric variations revealed that they were inconsistent with atmospheric phase variations due to the planet based on their phase and amplitude, but they could possibly be attributed to stellar activity. Comparing the RV analysis with the phase-folded photometric fluxes (see Figure~\ref{fig:RvvsFlux})) revealed a strong correlation between the two datasets (see Figure~\ref{fig:Correlate}). 

With these results in mind, we addressed what this means for the interpretation of the RV modulation observed near 3.9 days, previously attributed to a planetary signal. Stellar activity is a possible culprit, but the spectroscopic emission lines of this star do not correlate well with the photometric modulations of this star. Therefore, there is a wide field of opportunity for this target to be analyzed further to determine the source of the discrepancy between the photometric and spectroscopic behavior.

These results indicate that BD-06~1339b is, in fact, a likely false positive whose signature was induced by the activity of the star. Follow-up observations could help to resolve the discrepancy between the photometric and spectroscopic data. In particular, understanding the nature of the discrepancy would benefit from additional precision photometry of the star to improve the characterization of the stellar variability, alongside simultaneous spectroscopic activity indicators \citep{diaz2018b} and an extended RV baseline. Overall, reanalysis of this systems emphasizes the greater importance of further verifying the nature of confirmed RV planets as new data becomes available---especially for those that are low in mass and of high interest to demographics and atmospheric studies. 


\section*{Acknowledgements}

The authors acknowledge support from NASA grant 80NSSC18K0544, funded through the Exoplanet Research Program (XRP). T.F. acknowledges support from the University of California President's Postdoctoral Fellowship Program. This research has made use of the NASA Exoplanet Archive, which is operated by the California Institute of Technology, under contract with the National Aeronautics and Space Administration under the Exoplanet Exploration Program. This paper includes data collected with the TESS mission, obtained from the MAST data archive at the Space Telescope Science Institute (STScI). Funding for the TESS mission is provided by the NASA Explorer Program. STScI is operated by the Association of Universities for Research in Astronomy, Inc., under NASA contract NAS 5–26555. All of the data presented in this paper were obtained from the Mikulski Archive for Space Telescopes (MAST). STScI is operated by the Association of Universities for Research in Astronomy, Inc., under NASA contract NAS5-26555. Support for MAST for non-HST data is provided by the NASA Office of Space Science via grant NNX13AC07G and by other grants and contracts. This research made use of Lightkurve, a Python package for Kepler and TESS data analysis \citep{Lightkurve2018}.

\facilities{TESS, HARPS, NASA Exoplanet Archive}

\software{Astropy \citep{Astropy_Collaboration13, Astropy_Collaboration18},
          Astroquery \citep{Ginsburg19},
          \texttt{GLS} \citep{zechmeister2009},
          Lightkurve \citep{Lightkurve2018},
          \texttt{RadVel} \citep{fulton2018a},
          \texttt{RVSearch} \citep{rosenthal2021},
          Matplotlib \citep{Hunter07},
          NumPy \citep{Harris20}, 
          SciPy \citep{Virtanen20}
          }




\end{document}